\documentstyle[aps,preprint]{revtex}
\begin{document}

\draft

\title{ Vertical pairing of identical particles suspended in the plasma sheath }

\author{V. Steinberg$^1$, R. S\"{u}tterlin, A. V. Ivlev, and G. Morfill}

\address{Max-Planck-Institut f\"ur Extraterrestrische Physik,D-85740 Garching, Germany,\\
          $^1$ also: Department of Physics of Complex Systems,\\
          Weizmann Institute of Science, Rehovot, 76100, Israel}
\date{\today}
\maketitle

\begin{abstract}
It is shown experimentally that vertical pairing of two identical
microspheres suspended in the sheath of a radio-frequency (rf)
discharge at low gas pressures (a few Pa), appears at a well
defined instability threshold of the rf power. The transition is
reversible, but with significant hysteresis on the second stage. A
simple model, which uses measured microsphere resonance
frequencies and takes into account besides Coulomb interaction
between negatively charged microspheres also their interaction
with positive ion wake charges, seems to explain the instability
threshold quite well.
\end{abstract}

\narrowtext

Recent interest in the properties of complex plasmas -- charged
monodisperse microparticles suspended in an electron-ion
environment, is partially due to the possibility to model
condensed matter phenomena on an ``atomic'' level
\cite{thomas,chu,haya}. The particles are charged negatively in a
radio-frequency (rf) discharge (up to $10^3-10^4$ elementary
charges) and levitate in the lower plasma sheath, where gravity can be
balanced by an inhomogeneous vertical electric field. The
particles repel each other via a screened Coulomb interaction.
Inside a confining horizontal potential they arrange themselves
either into an ordered structure -- the ``plasma crystal'', or
form disordered, liquid- or gas-like, states
\cite{thomas1,melzer,quinn}.

At low gas pressures one has typically only a very few layers (as
few as one) for these structures. Ion streaming motion in the
plasma sheath produces a nonequilibrium environment, which is
reflected in the properties of the complex plasma structures and
makes them rather different from the known equilibrium ones
\cite{vlad,goree}. One of the striking observations at
sufficiently low gas densities (such that the ion-neutral mean
free path is not small compared with a typical distance between
microspheres) is an unusual ``stacking'' of the particles such
that adjacent horizontal layers are located on top of each other
\cite{chu,melzer,taka}. This vertical ``polarization'' of the dusty plasma
was ascribed to the attractive forces which result from the
focusing of the ion flux under each particle (ion wake effect,
see, e.g., Ref.~\cite{schweigert}). Thus, the attractive
interaction between the horizontal layers in a plasma crystal is
asymmetric, such that the attractive force is communicated only
downstream in the direction of the ion flow \cite{schweigert}. The
same arguments were applied to predict a ``binding'' force for a
possible ``dust molecule'' formation \cite{resen}. This
theoretical prediction was verified recently by experimental
studies of the competition between repulsive and attractive forces
acting on two {\it different} dust particles levitated on {\it
different} levels in the plasma sheath\cite{melzer1}.

In this Letter we present experiments on a new instability which
is observed for two {\it identical} microspheres suspended
initially on the {\it same} level in the plasma sheath. In this
work we will report on the quantitative investigations involving
(for simplicity) only two microspheres, which were undertaken to
clarify the basic behavior for many identical particles. The
instability appears first as a continuous (forward) bifurcation to
a state of the vertically separated microspheres, and then to a
discontinuous vertical pairing. Depending on gas pressure, the
last stage -- the pairing -- can be either strongly hysteretic
(for lower pressure) or weakly hysteretic (for higher pressure).

The experiments were performed in a standard Gaseous Electronics
Conference (GEC) rf Reference Cell \cite{GEC} with the lower
electrode powered at 13.56~MHz and the upper electrode grounded
(see Fig.~\ref{f1}). Argon gas at various pressures between 1 and
7 Pa was used, and the discharge power (or a rf peak-to-peak
voltage, $U_{\rm pp}$) was the control parameter of the
instability described. The electron temperature and density were
measured at the center of the discharge with a rf-compensated
passive Langmuir probe. At these conditions the electron
temperature $T_e$ was found to be within 2-5~eV, while the
electron density $n_e$ ranged from $10^7$ to $8\times
10^8$~cm$^{-3}$, so that the electron Debye length varied from
$\lambda_{{\rm D}e}\simeq 0.5$~mm (high $U_{\rm pp}$) to
$\lambda_{{\rm D}e}\simeq 5$~mm (low $U_{\rm pp}$). These values
of the Debye length are comparable to, or larger than the
separation distances measured, and thus Debye shielding will play
only a minor role in this rather low-density plasma. The
microspheres, suspended in the plasma, were illuminated by a laser
sheet of about $100~\mu$m thickness, and their imaging was
performed by external CCD cameras from the top and from the side
(in both cases via 300~mm objectives). We used polystyrene
(density $1.05$~g/cm$^3$) microspheres  with diameter
$7.6\pm0.1~\mu$m and mass $M\simeq2.4\times10^{-10}$~g. In order
to confine the microspheres horizontally we placed a quartz glass
cylinder of 20~mm height and 50~mm diameter on the lower electrode
(see Fig.~\ref{f1}). In contrast to a copper ring normally used to
confine particles, the glass cylinder produces extremely flat,
almost ``square well'' confining potential. The latter conclusion
is based on our observation of unusually flat shape of 2D plasma
crystal, particularly up to its edges, contrary to the trap formed
by a standard metal ring. The ``square well'' confining potential
provides us the opportunity to study two identical microspheres
suspended in the plasma sheath on the same initial height.

In order to quantitatively investigate the pairing instability and
to verify that it is indeed symmetry breaking, we conducted
experiments with two identical microspheres at different
pressures. Figure~\ref{f2} shows, how, at a pressure of 2 Pa, the relative position of one
microsphere, which eventually moves below and then under the other
(i.e., downstream) with respect to the eventual upstream
microsphere, varies as the power (or $U_{\rm pp}$) decreases. As
long as $U_{\rm pp}$ exceeds the threshold value $U_{\rm pp}^{\rm
th}\simeq55$~V, both particles are located on the same level
(initial position 1). When $U_{\rm pp}$ decreases below the
threshold, vertical separation starts (position 2) and grows
continuously up to position 3. A further small $U_{\rm pp}$
decrease causes the lower particle to ``jump'' to a position directly beneath the
upper one and to create a vertical pair (final position 4). By
reversing the process, i.e., increasing $U_{\rm pp}$, the
particles revert back to their initial horizontal configuration.
The transition is strongly hysteretic with respect to $U_{\rm
pp}$. Further measurements showed that the threshold, $U_{\rm
pp}^{\rm th}$, rapidly increases with pressure, while the relative
width of the hysteresis decreases with pressure.

Figure~\ref{f3} represents the forward (positions 1-5) and the
reverse (positions 6,7) transitions at $p=7$~Pa. The instability
(vertical separation) starts at $U_{\rm pp}^{\rm th}\simeq200$~V
and develops continuously until position 4. Then a small $U_{\rm
pp}$ decrease is accompanied by the discontinuous vertical pairing
(position 5). When the voltage is increased back, the particles
remain vertically paired until position 6 is reached, and then the
lower particle ``jumps'' to position 7. In Fig.~\ref{f4} the
vertical separation distance is plotted against the control
parameter, $U_{\rm pp}^{\rm th}-U_{\rm pp}$, at $p=7$ Pa. As in
the low pressure case, the transition is divided into two stages:
(i) continuous transition from the horizontal configuration, and
(ii) discontinuous, hysteretic transition to the final vertical
pairing. It is worth noting that at the onset of the continuous
transition strong {\it symmetrical} fluctuations of the vertical
particle separation are observed in the experiments. This means
that the pairing instability may be initiated with equal
probability by either microsphere in the pair. Therefore, this
transition is related to the symmetry breaking.

In the experiments both $U_{\rm pp}$ (at the fixed $p$) and $p$
(at the fixed $U_{\rm pp}$) were used as the control parameters of
the instability. But in a subsequent analysis, presented below, it
is much more convenient to use the resonance frequencies of
vertical, $\omega_z$, and horizontal, $\omega_r$, oscillations of
a single particle in the sheath as the control parameters. The
resonance frequencies are certain functions of $U_{\rm pp}$ and
$p$, and can be determined experimentally\cite{melzer2}.

We first examine the simplest model without wake effects. The model uses the
measured values of $\omega_z$ and $\omega_r$ as the control
parameters and provides an order of magnitude estimate for the
instability threshold. We consider a pair of two identical
particles of mass $M$ and (negative) charge $-Q$,
electrostatically confined in the harmonic potential well. If the
particles are separated horizontally by $R$ and vertically by
$\delta$ (i.e., each particle is displaced at $R/2$ and $\delta/2$
from the center), then their energy in the confinement is
$\frac14M(\omega_z^2 \delta^2+\omega_r^2R^2)$. Since the Debye
length is relatively large, the bare coupling energy of two unshielded
particles, $Q^2(R^2+\delta^2)^{-1/2}$, should be added to the
confinement energy. This system has two stable configurations --
horizontally (when $\omega_z>\omega_r$) or vertically (when
$\omega_z< \omega_r$) aligned at distances of $R=(2Q^2/M\omega_r^2
)^{1/3}$ or $\delta=(2Q^2/M\omega_z^2)^{1/3}$, respectively.

Our model is characterized in terms of the measured resonance
frequencies. Thus, it is necessary to trace these frequencies as
the control parameter, $U_{\rm pp}^{\rm th}-U_{\rm pp}$, is
varied. The typical dependence of $\omega_r$ and $\omega_z$ on the
value of $U_{\rm pp}$ at $p=$ 1 Pa is shown in Fig.~\ref{f5}. One
can see that the frequencies converge rapidly as $U_{\rm pp}$
decreases. Comparison of Figs.~\ref{f4} and \ref{f5} shows (albeit for different pressures) that
the transition starts {\it continuously} when $\omega_z$ still
{\it considerably exceeds} $\omega_r$. This implies that the
harmonic approximation with the isotropic particle field is too
simple. Thus, we have to include the wake effect in its simplest form.
For consistency in the nonlinearity analysis, we must also take
into consideration the anharmonicity of the
potential well in the vertical direction \cite{nonlinear}. The
corresponding expression for the total potential energy of a
particle pair in a 2D confinement is
\begin{equation}\label{1}
{\cal W}_{\rm pair}\simeq\frac{M}{4}\left(\omega_r^2R^2+\omega_z^2
\delta^2+\frac{\beta^*}{8}\delta^4\right)+\frac{Q^2}{\sqrt{R^2+
\delta^2}}-\frac{Qq}{2\sqrt{R^2+(\Delta+\delta)^2}}-\frac{Qq}{2
\sqrt{R^2+(\Delta-\delta)^2}}.
\end{equation}
In writing Eq. (\ref{1}) the simplest two-microsphere model of the wake is used
(see Fig.~\ref{f6}). We treat the excessive positive charge of the
wake, $q$, as point-like, located under the particle at distance
$\Delta$. We assume also that since both the particle and the wake
charges vary only slightly with variation in $R$ and $\delta$, so
that $Q$ and $q$ are approximately constant. The last two terms in
Eq. (\ref{1}) represent the particle-wake interaction, and
cross-terms of the particle-wake and wake-wake ($qq$) interactions
are neglected \cite{melzer1}. Since the coefficients multiplying the terms $\delta^3$ cancel
out for a pair of identical particles, the lowest order vertical anharmonicity in
Eq. (\ref{1}) is the fourth order term
with coefficient $\beta^*>0$ but whose value is here unknown, since it was not feasible
to measure it by technique used elsewhere \cite{nonlinear}.

The dependence $R(\delta)$ is determined from the equilibrium
condition in the radial direction, $\partial{\cal W}_{\rm pair}/
\partial R=0$:
\begin{equation}\label{2}
\frac{2Q^2}{(R^2+\delta^2)^{3/2}}-Qq\left(\frac1{[R^2+(\Delta+
\delta)^2]^{3/2}}+\frac1{[R^2+(\Delta-\delta)^2]^{3/2}}\right)=M\omega_r^2.
\end{equation}
The equilibrium value of $\delta$ is given by the condition
$\partial{\cal W}_{\rm pair}/\partial\delta=0$. A linear
combination of this condition and Eq. (\ref{2}) gives
\begin{equation}\label{4}
M({\omega_z}^2-{\omega_r}^2)\delta+\frac{M\beta^*}{4}\delta^3+Qq
\Delta\left(\frac1{[R^2+(\Delta+\delta)^2]^{3/2}}-\frac1{[R^2+
(\Delta-\delta)^2]^{3/2}}\right)=0.
\end{equation}
Expanding Eq.(\ref{4}) to third power in $\delta$ and substituting
$R(\delta)$ from Eq. (\ref{2}) provide a typical stationary
equation for the vertical displacement as the order parameter:
\begin{equation}\label{5}
\left(\frac{\omega_z^2}{\omega_r^2}-\Omega^2\right)\tilde\delta+
\frac{\beta^*R_0^2}{4\omega_r^2}\tilde\delta^3-\frac{35(\Omega^2-1)
\widetilde\Delta^2}{6(1+\widetilde\Delta^2)^2}\left(\frac{1+\frac87
\frac{q/Q}{(1+\widetilde\Delta^2)^{5/2}}}{1-\frac{q/Q}{(1+
\widetilde\Delta^2)^{5/2}}}\right)\tilde\delta^3=0.
\end{equation}
Here we have introduced the dimensionless parameters, $\tilde\delta=
\delta/R_0$ and $\widetilde\Delta=\Delta/R_0$ [normalized by
$R_0=R|_{\delta=0}$ from Eq. (\ref{2})], and the critical
frequency ratio,
\begin{equation}\label{3}
\Omega\equiv\sqrt{1+\frac{6Qq\widetilde\Delta^2}{M\omega_r^2 R_0^3
(1+\widetilde\Delta^2)^{5/2}}}.
\end{equation}
The instability criterion $(\omega_z/\omega_r)_{\rm cr}=\Omega$ is
obtained from Eq. (\ref{5}) by setting the coefficient of
$\tilde\delta$ equal to zero. At $\omega_z/\omega_r>\Omega$, the
energy ${\cal W}_{\rm pair}(\delta)$ has a minimum at $\delta=0$
and the equilibrium configuration is horizontal. In the opposite
case, $\omega_z/\omega_r<\Omega$, the $\delta=0$ state becomes
unstable, and the particles start separating vertically. From
Fig.~\ref{f5} we note that the separation starts at
$(\omega_z/\omega_r)_{\rm cr}\simeq4$ (when $U_{\rm pp}\lesssim
U_{\rm pp}^{\rm th}$). Assuming $Q\sim q$ and
substituting Eq. (\ref{2}) at $\delta=0$ into Eq. (\ref{3}),
one gets $\Omega$ as a function of $\widetilde\Delta$ only. Then at
$\widetilde\Delta\sim 1$  we
obtain $\Omega\sim3$. Thus, the threshold of the pairing
instability is adequately described by the model. A saturation of
the order parameter, $\tilde\delta$, above the instability
threshold in the first continuous stage of the transition (see
Fig.~\ref{f4}) could be reached if the sum of both coefficients of
$\tilde\delta^3$ in Eq. (\ref{5}) is positive. By using the expression
$\Omega(\widetilde\Delta)$  and the value of the ratio $Q/q\sim 1$
we achieve this condition at $\beta^*R_0^2/\omega_r^2\gtrsim 7$, or $\beta^*/
(2\pi)^2\gtrsim 20$~Hz$^2$/mm$^2$. Our recent experiments
\cite{nonlinear} indicate that at pressures below $\sim10$~Pa the
anharmonic coefficients become relatively large: For example,
$\beta/(2\pi)^2 \simeq20 $~Hz$^2$/mm$^2$ at $p\sim1$~Pa and
$U_{\rm pp}\simeq 70$~V. At smaller $U_{\rm pp}$, the whole
vertical structure of the sheath changes dramatically, because the
Debye length increases by the order of magnitude. Hence, the
anharmonic coefficients should also increase rapidly as $U_{\rm
pp}$ decreases, and it is plausible that $\beta^*$ will provide the instability
saturation. Then, as follows from Eq.(\ref{5}), $\tilde\delta
\propto\sqrt{\Omega-\omega_z/\omega_r}$ close to the threshold.
However, failing an accurate experimental determination of the shape of the sheath potential
for the actual conditions of the experiment, further testing of the model via the
experimental value of $\delta$ is not feasible here.

The modelling of the second (discontinuous) stage of the
instability is rather complicated problem: When the lower particle
approaches the wake, its actual ``shape'' becomes crucial, and
the approximation of the point-like wake charge cannot be used.
This very nonlinear second stage will almost certainly involve nonlinear wake dynamics,
and these can probably only be adequately treated by three-dimensional numerical simulations.
 We would like to point out that the effect of the wake is stronger
at lower $p$ and $U_{\rm pp}$, when $\lambda_{{\rm D}e}$ is larger.
At the same time, the ion mean free path should exceed
$\lambda_{{\rm D}e}$ (otherwise, the wake effect weakens due to
the ion scattering on neutrals). Both these requirements are well
satisfied in our experiments.

V. S. is grateful for the support of the Alexander von Humboldt
Foundation, Germany, during his stay at MPE. Many useful critical
remarks of an anonymous Referee are gratefully acknowledged.

\begin{figure}
\caption{Experimental setup using a quartz glass cylinder as a
confining electrode.} \label{f1}
\end{figure}

\begin{figure}
\caption { Sequence of the relative particle positions during the
pairing instability as  $U_{\rm pp}$,
decreases (pressure $p=2$~Pa). The steps are: (1) $U_{\rm pp}\geq
U_{\rm pp}^{\rm th}\simeq55$~V -- horizontal configuration, (2)
$U_{\rm pp}\simeq45$~V and (3) $U_{\rm pp}\simeq37$~V --
continuous separation, (4) $U_{\rm pp}\simeq36$~V -- discontinuous
vertical pairing.} \label{f2}
\end{figure}

\begin{figure}
\caption{Sequence of the relative particle positions during the
pairing instability as  $U_{\rm pp}$,
decreases (1-5) and increases back (6,7) (pressure $p=7$~Pa). For
decreasing voltage the steps are: (1) $U_{\rm pp}\geq U_{\rm
pp}^{\rm th}\simeq200$~V -- horizontal configuration, (2) $U_{\rm
pp}\simeq110$~V, (3) $U_{\rm pp}\simeq85$~V, and (4) $U_{\rm
pp}\simeq78$~V -- continuous separation, (5) $U_{\rm pp}\simeq
76$~V -- discontinuous vertical pairing; For increasing voltage
the steps are: (6) $U_{\rm pp}\simeq93$~V -- continuous vertical
movement and (7) $U_{\rm pp}\simeq95$~V -- discontinuous
reversibility of the pair configuration.} \label{f3}
\end{figure}

\begin{figure}
\caption{Vertical separation of a particle pair vs. the control
parameter, $U_{\rm pp}^{\rm th}-U_{\rm pp}$, for two stages of the
pairing transition at $p=7$~Pa. The hysteretic nature of the
second stage is obvious from the plot. The arrows denote the
direction of the variation observed.} \label{f4}
\end{figure}

\begin{figure}
\caption{Horizontal, $\omega_r$ ($\diamond$), and vertical,
$\omega_z$ (${\scriptsize +}$), resonance frequencies of a single
particle as function of $U_{\rm pp}$
(at a pressure $p=1$~Pa). The vertical dashed line shows the
threshold voltage $U_{\rm pp}^{\rm th}\simeq35$~V (onset of
the vertical separation).} \label{f5}
\end{figure}

\begin{figure}
\caption{ Schematics of the simplified model of the wake
potential. } \label{f6}
\end{figure}

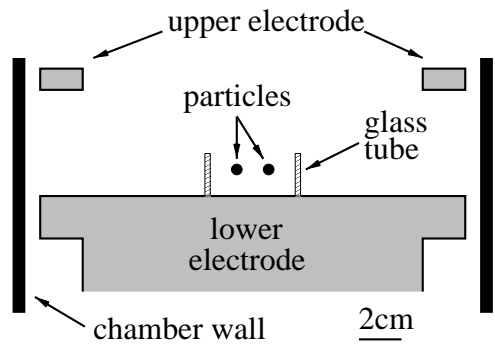
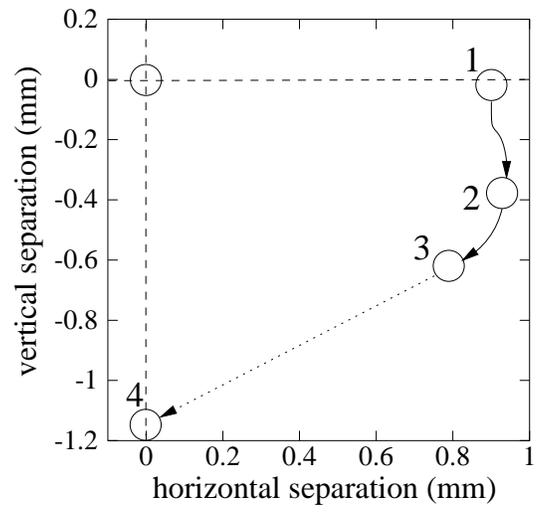
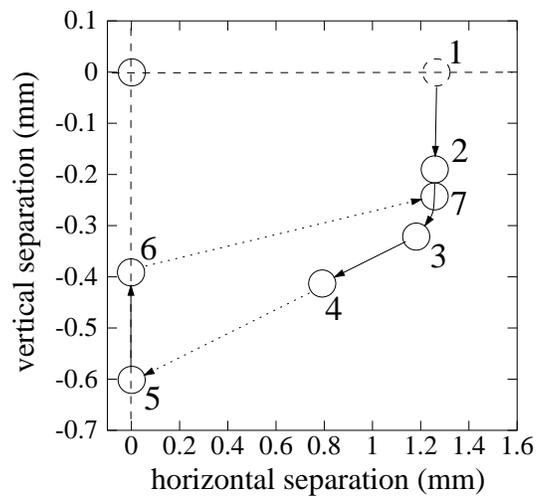
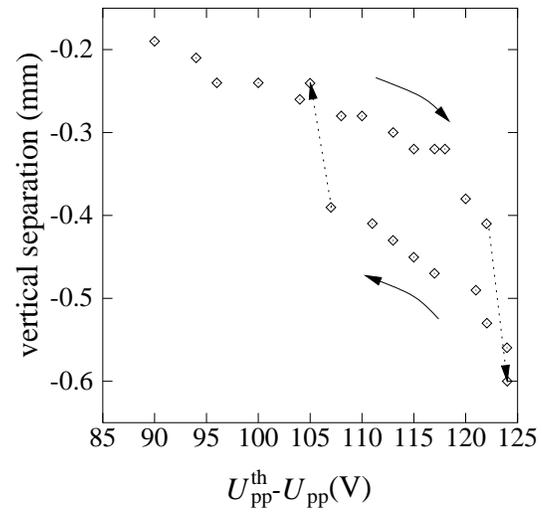
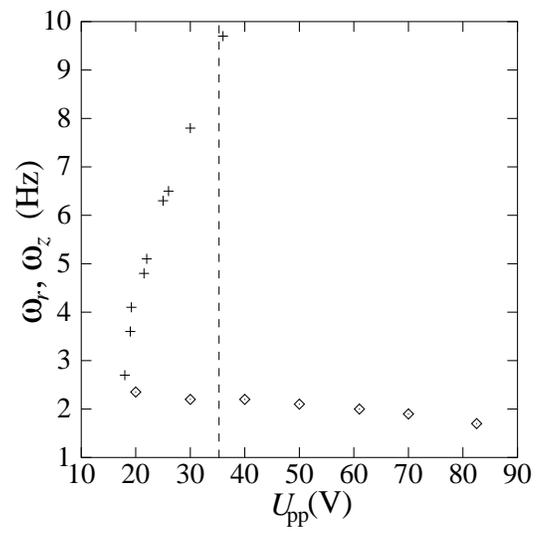
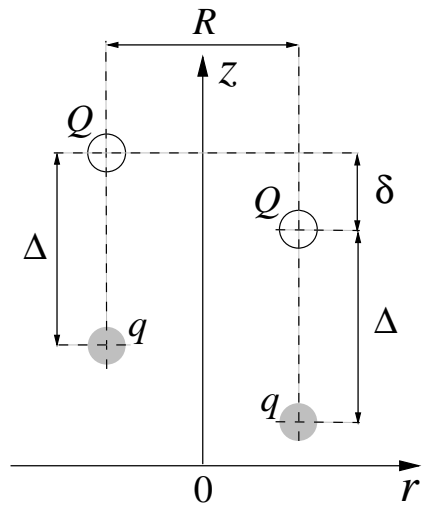
\end{document}